\documentclass[]{aa}
\usepackage{txfonts}
\usepackage{psfig,longtable,lscape}
\usepackage{graphicx}
\usepackage{rotating}

\begin{document}

\title{Rotation periods of Post-T Tauri stars in Lindroos systems
\thanks{Based on observations collected at Cerro Tololo Inter-American
Observatory and the European Southern 
Observatory, La Silla,  under project 66.C-0119(B)}}

\author{N. Hu\'elamo\inst{1}
\and M. Fern\'andez\inst{2}
\and R. Neuh\"auser\inst{3}
\and S.J. Wolk\inst{4} }

\offprints{Nuria Hu\'elamo,
\email{nhuelamo@eso.org}}

\institute{European Southern Observatory, Alonso de Cordova 3107, 
Casilla 19001, Santiago, Chile
\and Instituto de Astrof\'{\i}sica de Andaluc\'{\i}a, CSIC, 
Camino Bajo de Hu\'etor 24,  E-18080 Granada, Spain
\and Astrophysikalishes Institut, Universit\"at Jena, Schillerg\"asschen 2-3, 
D-07745 Jena,  Germany 
\and Harvard Smithsonian Center for Astrophysics, Mail Stop 72, 
Cambridge, MA02138, USA}

\date{Received/Accepted}

\abstract{We present a rotational study of Post-T Tauri stars (PTTSs)
in Lindroos systems, defined as binaries with early type primaries on
the main-sequence (MS) and late-type secondaries on the
pre-main-sequence (PMS) phase.  The importance of this study in
comparison with previous ones is that the Lindroos sample is not X-ray
selected so we avoid a possible bias towards fast rotators. In
this preliminary study we have monitored eleven stars in the {\em
UBVRI} bands during two campaigns of ten consecutive nights each.
Eight of the observed PTTSs show periodic modulations in their
lightcurves and the derived periods range from 1.9\,d to 8.0\,d.  The
comparison of these results with theoretical rotational tracks based
on disk-star locking theory shows that star-disk decoupling times of
1-20\,Myr could reproduce the rotational properties of the targets,
assuming an initial rotation period of $\sim$ 8\,d and a mass of
1\,M$_{\odot}$.  We have studied the rotation-activity relations of
Lindroos PTTSs and compared them with those found in other groups of
PMS and zero-age main-sequence (ZAMS) $\sim$1\,M$_{\odot}$ stars. The
Lindroos sample displays activity-rotation relations very similar to
those found in TTSs.  It contains a mixture of very active stars, with
$L_{\rm X}/L_{\rm bol}$ ratios close to the saturation level of --3,
and less active (unsaturated) stars.  This could be the result of
different star-disk decoupling times. Future monitoring of a larger
and unbiased sample of PTTS will be important to confirm the
significance of these results.

 } \maketitle

\section{Introduction}\label{intro}

 In the last decade a large number of works have been devoted to study
the angular momentum evolution of T Tauri stars (TTSs), that is,
late-type stars contracting towards the main-sequence (MS).
Observations of late-type pre-main sequence (PMS) stars have allowed
investigators to derive the rotational properties of TTSs in different
star forming regions (e.g. Bouvier et al. 1993, Edwards et al. 1993,
Choi \& Herbst 1996, Wichmann et al. 1998, Herbst et al. 2001a, Rebull
2001, Lamm et al. 2004).  The earliest studies reported the existence
of a bimodal distribution of rotational periods among TTSs in Taurus:
while accreting TTSs surrounded by disks were generally slow rotators
with periods longer than 4.4 days, non-accreting TTSs were faster
rotators showing periods that range from a few hours to 3 days. Recent
rotational studies on late-type PMS stars in Orion are more
controversial: while some works do not find a bimodal distribution
among late-type PMS stars (e.g. Stassun et al. 1999), other studies
(e.g. Herbst et al. 2001b) have shown that the bimodal distribution is
present but is mass-dependent, that is, it is only found in stars with
masses above a certain limit ($M\geq$0.25\,$M_{\odot}$).

Most of the results from the rotational studies support disk-locking
as the most plausible mechanism to regulate the angular momentum of
late-type PMS stars (see Rebull et al. 2002). The so-called
disk-locking theory (e.g. K\"onigl 1991, Collier Cameron \& Campbell
1993, Shu et al. 1994, Bouvier et al. 1997a) explains the rotational
properties of late-type PMS stars as a result of magnetic coupling
between the star-disk system, that prevents the star from spinning-up
during the mass accretion phase. Once the disk is dissipated the
couping no longer exists and the star begins to rotate faster due to
on-going contraction to the MS.

 The rotational properties of late-type Zero-Age Main-Sequence (ZAMS)
 dwarfs in open clusters like the Pleiades or $\alpha$ Per have been
 also extensively studied in different works (e.g. Stauffer et
 al. 1989, Soderblom et al., 1993, Prosser et al. 1995, Allain et al.
 1996, Queloz et al. 1998, Terndrup et al. 2000).  These works have
 reported a large scatter in the rotational rates of these stars:
 while half of the objects display projected velocities ($v.\sin i$)
 of a few km/s, half of them are faster rotators with velocities up to
 200~km/s (the so-called ultra fast rotators, UFR's).  A decoupling
 between the radiative core and the convective envelope has been
 studied to explain the large spread of rotational properties among
 ZAMS stars (e.g. Allain 1998).  In the case of older clusters like
 the Hyades (Radick et al. 1987, Stauffer et al., 1997), the rotation
 rates are significantly smaller and most of the late-type stars
 display projected velocities of $v.\sin i < 10$ km/s.
 
Post-T Tauri stars (PTTSs, Herbig 1978) are intermediate between the
oldest T Tauri stars and the youngest dwarfs in open clusters.  As
noted by Bouvier et al. (1997b), it is precisely at this evolutionary
stage when the stellar interior changes from completely convective to
radiative, so the internal changes may be reflected in the surface
rotation properties.  Two main studies have tried to derive the
rotational properties of PTTSs: Bouvier et al. (1997b) and Wichmann et
al. (1998) in the Taurus and Lupus star forming regions (SFR),
respectively.  Both works showed that most of the PTTSs under study
were fast rotators. However, this conclusion could be the result of
their sample selection criteria which included the X-ray detection of
the sources.  X-ray selected samples tend to be biased towards fast
rotators: late-type stars show a connection between magnetic activity
and rotation, with the strongest X-ray emitters being the fastest rotators
(e.g. Bouvier 1990, Neuh\"auser et al. 1995). Hence, an unbiased
sample of PTTSs is required to understand the rotational properties of
late-type stars at that evolutionary stage.

In this context, the so-called Lindroos sample of PTTSs (Lindroos 1986) 
is important because it is not X-ray selected and complements
the previous samples of PTTSs. Lindroos systems are defined as visual
binary systems mainly comprised of early-type primaries on the MS and
late-type secondaries (Lindroos 1985). The ages of Lindroos primaries
have been derived through photometric and spectroscopic observations,
showing values between 10$^7$-10$^8$ yr. This age interval is
comparable to the contraction time-scale of late-type stars to the MS.
If the systems are physically bound, the secondaries are late-type PMS
stars still contracting to the ZAMS, that is, PTTSs.

Lindroos (1985, 1986) carried out an exhaustive study of 253 visual
binary systems and found 78 binary or multiple systems likely to be
physical. A total of 45 companion candidates with spectral types later
than F0 were selected among the binary and multiple systems.
Pallavicini et al. (1992) and Mart\'{\i}n et al. (1992) have reported
the presence of indicators of youth (Li {\sc I} absorption line) and
activity (H$\alpha$ emission line) in 40\%~of the Lindroos late-type
secondaries.  The analysis of the X-ray emission from 22 Lindroos
systems resolved by the ROSAT High Resolution Imager (HRI) has shown
that most of the Lindroos PTTSs candidates are X-ray emitters, with
X-ray luminosities comparable to those from younger late-type PMS
stars (Hu\'elamo et al. 2000).

In this preliminary work we have analyzed the rotational
properties of bona-fide PTTSs in Lindroos systems.  Given that the
sample is not X-ray selected, it is ideal to study the rotational
properties of PMS stars in that evolutionary phase. Our main goal is
twofold: first, we will compare the observed rotational properties
with theoretical predictions of angular momentum evolution of
late-type stars and, second, we will derive the rotation-activity
relations for our sample and compare them with groups of late-type
stars at different evolutionary stages.  This way, we can trace the
evolution of activity parameters as stars evolve and reach the MS.
The Lindroos sample is described in Section~2.  The optical
observations of the sample are described in Section~3, while the
analysis of the data is provided in Section~4. The main results are
described in sections~5, 6 and 7, while our preliminary
conclusions are drawn in Section~8.

   \begin{table*}[htbp]
      \begin{center}
      \caption[table1]{Stellar properties of the selected sample of Lindroos PTTSs}\label{table1}
        \begin{tabular}{llllrclll}
          \noalign{\smallskip}
          \hline
          \noalign{\smallskip}

HD  & Sp.Type$^1$ & Distance$^2$ & Sep.$^3$ & V$^4$ & A$_{\rm v}$$^5$ & 
$log\,L_{\rm X}$$^6$ & Mass$^7$ & Age$^8$ \\
&   &  (pc)       & (AU) & (mag) & (mag) & (erg/s) & (M$_{\rm \odot}$) & Myr\\
          \noalign{\smallskip} 
          \hline 
          \noalign{\smallskip}
560\,B    & G5 + B9  & 100$\pm$9  & 770  & 10.37 & 0.00 & 30.56    & 0.95 & 20/30/36/30\\
17543\,B  & F8 + B6  & 187$\pm$37 & 4625 & 10.73 & 0.21 & 30.17    & 1.18 & 24/28/-/25\\
23793\,B  & F3 + B3  & 173$\pm$31 & 1557 & 9.41  & 0.08 & 30.25   & -    & $<$28$^*$\\
27638\,B  & G2 + B9  &  82$\pm$8  & 1558 & 8.43  & 0.00 &$<$28.49 &1.25  & 12/15/19/15\\
33802\,B  & G8 + B8  &  74$\pm$4  & 940  & 9.92  & 0.01 & 30.03   & 0.90 & 25/50/47/37 \\ 
38622\,C  & G2 + B2  & 245$\pm$54 & 6125 & 12.01 & 0.04 & 29.99   & 1.05 & 30/100/-/110\\ 
40494\,B  & G8 + B3  & 263$\pm$38 & 8889 & 12.66 & 0.00 &$<$28.57 & 0.94 & 35/100/-/76 \\ 
53191\,B  & G3 + A0  & 207$\pm$25 & 3519 & 11.75 & 0.05 & 29.53   & 1.00 &  30/80/-/39\\ 
60102\,B  & G8 + B9.5& 206$\pm$25 & 3378 & 11.86 & 0.22 & 30.62   & 1.00 &  15/25/-/20\\ 
86388\,B  & F5 + B9  & 177$\pm$21 & 1628 & 9.98  & 0.02 & 29.42   & 1.28 &  -/30/-/110\\ 
90972\,B  & F9 + B9.5& 147$\pm$16 & 1617 & 9.65  & 0.03 & 30.11   & 1.25 &  14/18/18/15\\ 
108767\,B & K2 + B9  &  27$\pm$9  & 648  & 8.43  &  0.00& 28.95   & 0.8  &  40/90/-/93 \\ 
109573\,B$^+$ & M2.5 + A0 & 67$\pm$3& 509& 13.3  & 0.00 & 29.75   & 0.3  & 10 \\ 
113703\,B & K0 + B4  & 127$\pm$12 & 1448 & 10.8  & 0.00 & 30.22   & 0.98 & 15/25/19/23 \\
113791\,B & F7 + B2  & 126$\pm$13 & 3163 & 9.38  & 0.00   & 29.98   & 1.22 & 18/22/11/21 \\
127304\,B & K1 + A0  & 106$\pm$8  & 2734&  11.37 & 0.04 & $<$28.73& 0.84 &50/100/-/100 \\
129791\,B & K5 + B9.5& 129$\pm$16 & 4554 & 12.93 & 0.26 & 29.82   & -    & 39$^*$ \\
143939\,B & K3 + B9  & 167$\pm$27 & 1436 & 11.80 & 0.00 & 30.52   & 1.15 & 2.5/7/8.3/6 \\
        \noalign{\smallskip}
        \hline \\
        \end{tabular}
\end{center}
{\bf Notes:} 1. Adopted from Pallavicini et al. (1992). The spectral
type of the early-type primary star is also provided; 2. Deduced from
Hipparcos parallax of the primary star; 3. Separation between the
stars of the Lindroos binary system using Hipparcos parallax;
4. Adopted from Pallavicini et al. (1992) 5. Adopted from L86; 6.
Adopted from Hu\'elamo et al. (2000); 7.  Masses estimated adopted
from Gerbaldi et al. (2001) using Palla \& Stahler (1999) evolutionary
tracks; 8.  Ages adopted from Gerbaldi at al. (2001). The four entries
correspond to the ages computed with D' Antona \& Mazzitelli (1998),
Palla \& Stahler (1999), Siess et al. (2000) and Tout et al. (1999)
evolutionary tracks, respectively; $^+$ Data from Webb et al. (1999);
$^*$ Ages adopted from L86; $-$ data not available.

\end{table*}

\section{The Lindroos sample of PTTSs}

The Post-T Tauri stars in the sample were selected according to these
criteria: their ages must be within 10-100\,Myr and they must show
indications of being physically bound to their primaries.  The ages of
the Lindroos PTTSs have been derived using four sets of evolutionary
tracks (Gerbaldi et al. 2001).  Different models provide different
age estimations (see Table~1) but, in general, all lie within the selected
interval. Most of the late-type stars in the sample show ages and/or
radial velocity measurements consistent with those of the early-type
primaries. Moreover, if the systems are bound the secondaries are in
the PMS phase.  Hence, we have also looked for indicators of youth
among the sample to reinforce the evidences of being physically bound
to their primaries (see Jensen 2002).  All these properties are
summarized in Table~7 of Hu\'elamo et al. (2000). The final sample
contains 18 stars.

The Lindroos sample is not X-ray selected implying that our rotational
study should not be biased towards fast rotators.  However, Lindroos
PTTSs are found in binary systems and we were concerned 
that such binarity could introduce a different bias in our analysis.
We have analyzed whether the Lindroos secondary stars studied here are
representative PTTSs or, on the other hand, they are somehow
"special" PTTSs because of being members of binary systems.  In the
next subsections we have analyzed two binarity effects that can
influence the rotational evolution of these young objects: first, we
have analyzed if the separation between the members of the binary
systems is small enough to produce a circularization of their orbits
and, subsequently, a synchronization of both the orbital and rotation
period. Secondly, we have analyzed the effect of the UV radiation
field of the early-type primaries on the late-type secondaries.  Given
that such radiation can accelerate the dissipation of circumstellar
matter around PMS stars, it could produce an earlier spin-up of the
PTTSs (in the framework of the disk-star locking theory).

\subsection{Binarity and rotation: Circularization}

It is an observational fact that close binary systems with orbital
periods smaller than a given critical value (called the {\em
circularization period}) evolve to circularize their orbits
({\em e}\,=\,0). In this process, a synchronization of their
rotational and orbital periods is observed (e.g. Zahn 1977, Mathieu
1992, Melo et al. 2001). The mechanism that produces such
circularization is still not well understood (i.e. a tidal-torque
mechanism or a hydrodynamical mechanism).

The circularization is a slow process, so that the larger the
separation of the system the longer it takes to circularize the orbit.
Observations of PMS binary systems have shown that all short period
($P \leq 7.56^d$) binaries display circular orbits at ages of 10$^6$
yrs (Melo et al. 2001). This period can be taken as a circularization
cutoff period for PMS stars.  The lack of circular or near-circular
({\em e} $<$0.1) orbits for PMS binaries with orbital periods longer
than the cutoff period is evident for systems with $P\geq1000^d$
(Duquennoy \& Mayor 1991, Pan et al. 1998).  Taking into account the
separations and the masses of the Lindroos systems, we have derive
orbital periods larger than 1000\,yr.  Hence, we do not expect any
circularization process among the Lindroos sample.

\subsection{Binarity and rotation: Stellar Winds}

B-type stars mostly radiate in the far ultraviolet (FUV) range.  If
FUV radiation impacts onto a circumstellar disk it can produce its
photoevaporation whenever the incident FUV flux is of the order of
10$^4$ $\times$ $G_O$, being $G_O$ the average UV photon flux in the
local interstellar medium (Johnstone et al. 1998). As an example, such
a radiation flux is produced by O-type stars in the Orion Trapezium
which photoevaporate disks up to a distance of 1pc (St\"orzer \&
Hollenbach 1999). In the case of the Lindroos systems, the presence of
a B-type star close to a late-type companion could produce the same
effect.

As shown in Table~\ref{table1}, the spectral types of the Lindroos
primaries range between B2 and A0. As discussed by Bouvier \& Corporon
(2001), the distance up to which these stars produce the minimum flux
required to photoevaporate disks is of the order of 10$^4$\,AU for a
B3 star and about 500\,AU for a A0 star.  Most of the Lindroos
primaries are late B-type stars and the separations of the systems
range between 600\,AU and 9000\,AU (with a mean value of $\sim$
3000\,AU). Hence, we do not expect any effect of the UV radiation
field from the primaries into the secondaries. There are two
exceptions, HD\,113703\,B and HD\,113791\,B, with early B-type
primaries (B4 and B2) and separations smaller than 3500\,AU. These two
stars will be discussed in section~\ref{sec:rotat-prop-lindr}.

\section{Data acquisition}

Lindroos PTTSs are spread over different regions in the Northern and
Southern hemisphere. In order to monitor the stars in the northern
hemisphere, four weeks were allocated at the 1.5m Spanish telescope
and the 1.23m telescope at the Calar Alto Observatory (CAHA).
Unfortunately, we could not obtain good quality data due to poor
weather conditions.

We observed the southern sources in two campaigns in different
observatories: seven sources were observed at the 0.9m CTIO telescope
at the beginning of January 2001, while a second campaign at the 1.5m
Danish telescope in La Silla Observatory in February 2001 allowed us
to observe the rest of the targets.  In both campaigns (CTIO and La
Silla) we monitored the sources during 10 consecutive nights each.
The stellar data of the observed sample is summarized in
Table~\ref{table1}. Note that two of the stars in the sample could not
be monitored: HD108767\,B was too bright and saturated the detector,
while HD\,109573\,B, the only M-type star of the sample, is too faint
and too close to the early-type primary.

\subsection{CTIO 0.9m telescope data}

The observations at the 0.9\,m CTIO telescope were carried out during
ten consecutive nights in January 2001 (from the 4$^{\rm th}$ to the
13$^{\rm th}$).  The detector used was an ARCON
2024$\times$2024~pixels CCD camera, which is a multi-readout camera
with four different amplifiers.  The plate-scale of the camera is
0.396''/pixel, providing a total field of view of $13\arcsec \times
13\arcsec$.  We made use of the {\em UBV(RI)$_C$} filter set.

We observed the Lindroos secondary stars during ten nights with random
spacing times to avoid as much as possible false periodicities.  Most
of the stars were observed at least twice during the same night to be
sensitive to periods shorter than one day.  The exposure times of the
scientific images ranged between 0.5\,s and 10\,s. The shortest
exposure times were necessary to avoid the light of the primary star
in those binaries with smaller separations. In these cases, several
consecutive exposures were taken in each filter in order to average
them and to increase the signal to noise ratio (SNR) of the final
images.

The data were reduced using the QUADPROC/CCRED package into IRAF
(Image Reduction and Analysis Facility\footnote{IRAF is distributed by
the National Optical Astronomy Observatories, which is operated by the
Association of Universities for Research in Astronomy, Inc.  (AURA)
under cooperative agreement with the National Science Foundation.}),
specially designed to reduce multi-readout data.  All the images were
bias subtracted and flatfield corrected.  The shutter delay of the
camera is not negligible, so we have applied a shutter correction to
the images to compensate the inhomogeneous illumination of the
detector at short exposure times.  In order to do this we created
shutter masks at different exposures times.  After their
normalization, we corrected the scientific images multiplying by the
corresponding shutter mask.  The correction must be applied to images
with exposures times shorter than 10\,s.
 
We carried out aperture photometry on the sources.  For the closest
pairs (separations $<$ 10$\arcsec$) with a very bright primary
($V$-mag$<$6\,mag), the emission from the late-type secondary could be
contaminated with the light from the early-type star.  In these cases
we have applied multi-aperture photometry on the secondaries
correcting the magnitudes with field stars.  We have measured the
brightness of the objects in both the individual and the averaged
images to check the reliability of the combined image.  In general,
there is a good agreement between the magnitudes derived in this two
ways.  We have assumed that the uncertainties in the final magnitudes
are given by the ratio of the standard deviation of the mean of the
individual magnitudes to the square root of the number of individual
observations.

Apart from the Lindroos objects we have also selected objects in the
field of view as comparison stars.  We have used these stars to carry
out differential photometry on the Lindroos secondaries.

\subsection{La Silla 1.5m Danish telescope data}

The data from the 1.5m telescope were collected from the 29$^{\rm th}$
January to the 8$^{\rm th}$ February 2001.  The telescope is equipped
with DFOSC, a 2024$\times$2024~pixels CCD camera with a plate-scale of
0.39$\arcsec$/pixel.  The total field of view is
13$\arcmin\times$13$\arcmin$.  We used the {\it UBVRi} filter set.

The data reduction was carried out using the CCDRED package within
IRAF. The images were bias and flat-field corrected.  DFOSC is
equipped with a fast shutter (the shutter delay is of the order of
milliseconds), so in this case it was not necessary to apply a shutter
correction to the data.  For those binaries with smaller separations
we took several short exposures in each filter.  The final images are
the result of combining all individual frames.

The optical magnitudes have been measured following the same procedure
described for the CTIO data: we carried out aperture photometry on the
sources; for the closest pairs (separation $<$ 5$\arcsec$) we carried
out multi-aperture photometry correcting with field stars.  Finally,
we carried out differential photometry of all the targets using
comparison stars which were monitored during the ten nights.

\section{Analysis of the data}

\subsection{Variability analysis}

\begin{figure*}
 \centering
 \resizebox{18cm}{!}{\includegraphics{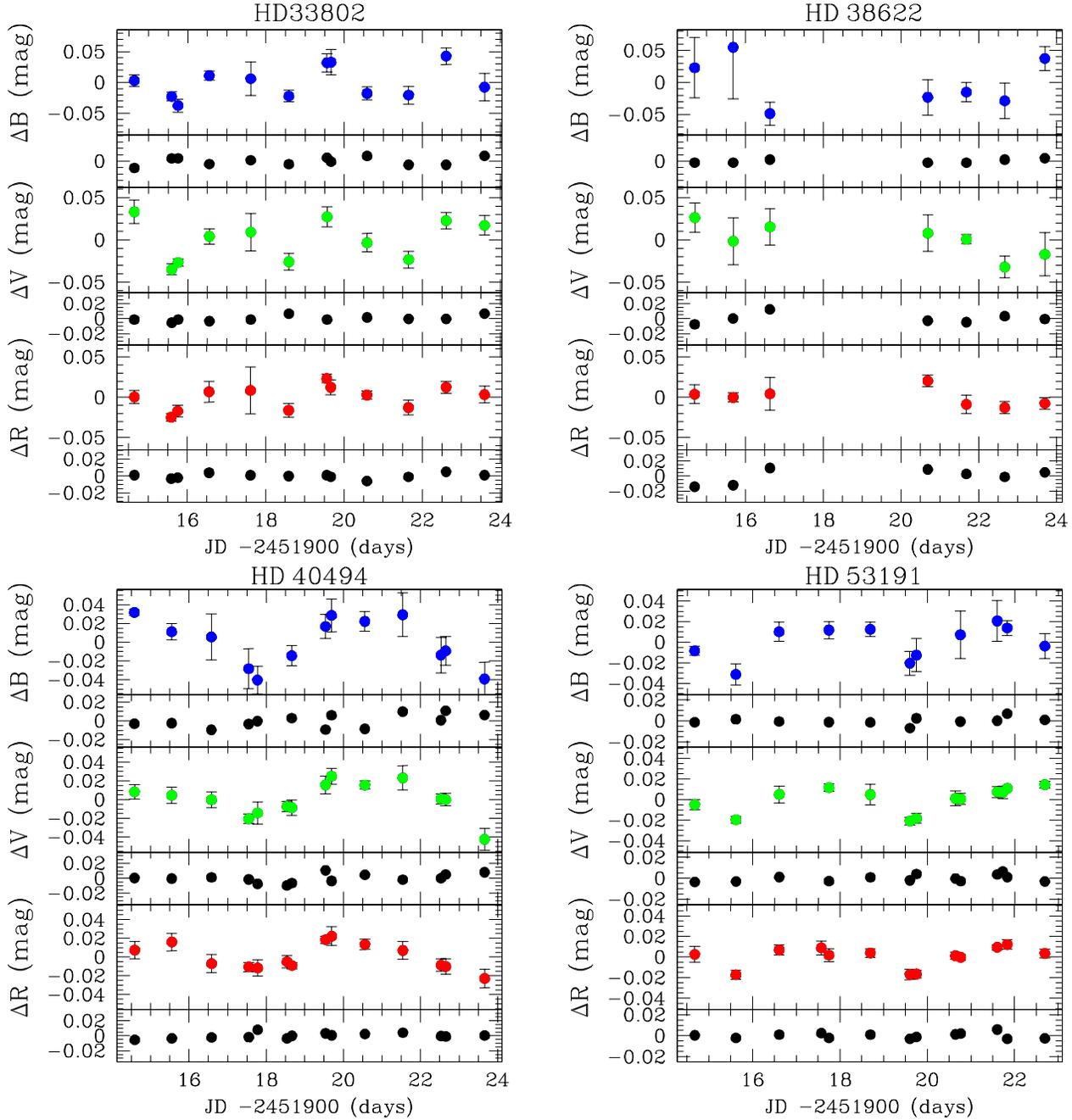}}
   \caption{{\em BVR} lightcurves of the observed Lindroos PTTSs. The
  differences in magnitudes between the target and the 'artificial
  comparison' star (see text) are displayed versus Julian Date at
  different filters for the variable stars. }
   \label{fig:light}
\end{figure*}
\addtocounter{figure}{-1}

\begin{figure*}
\resizebox{18cm}{!}{\includegraphics{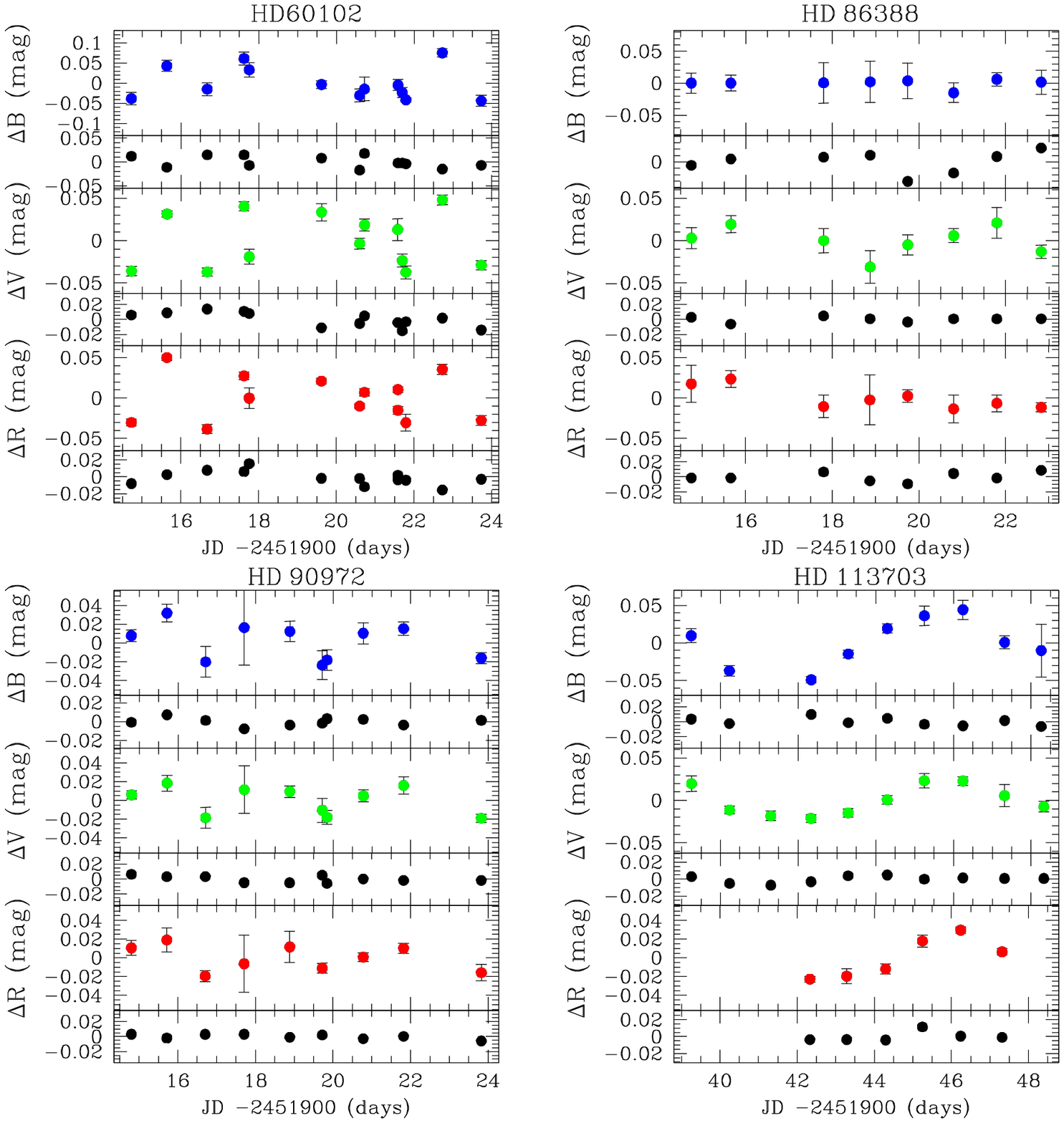}}
   \caption{cont.}
\end{figure*}

\addtocounter{figure}{-1}

\begin{figure*}
\resizebox{18cm}{!}{\includegraphics{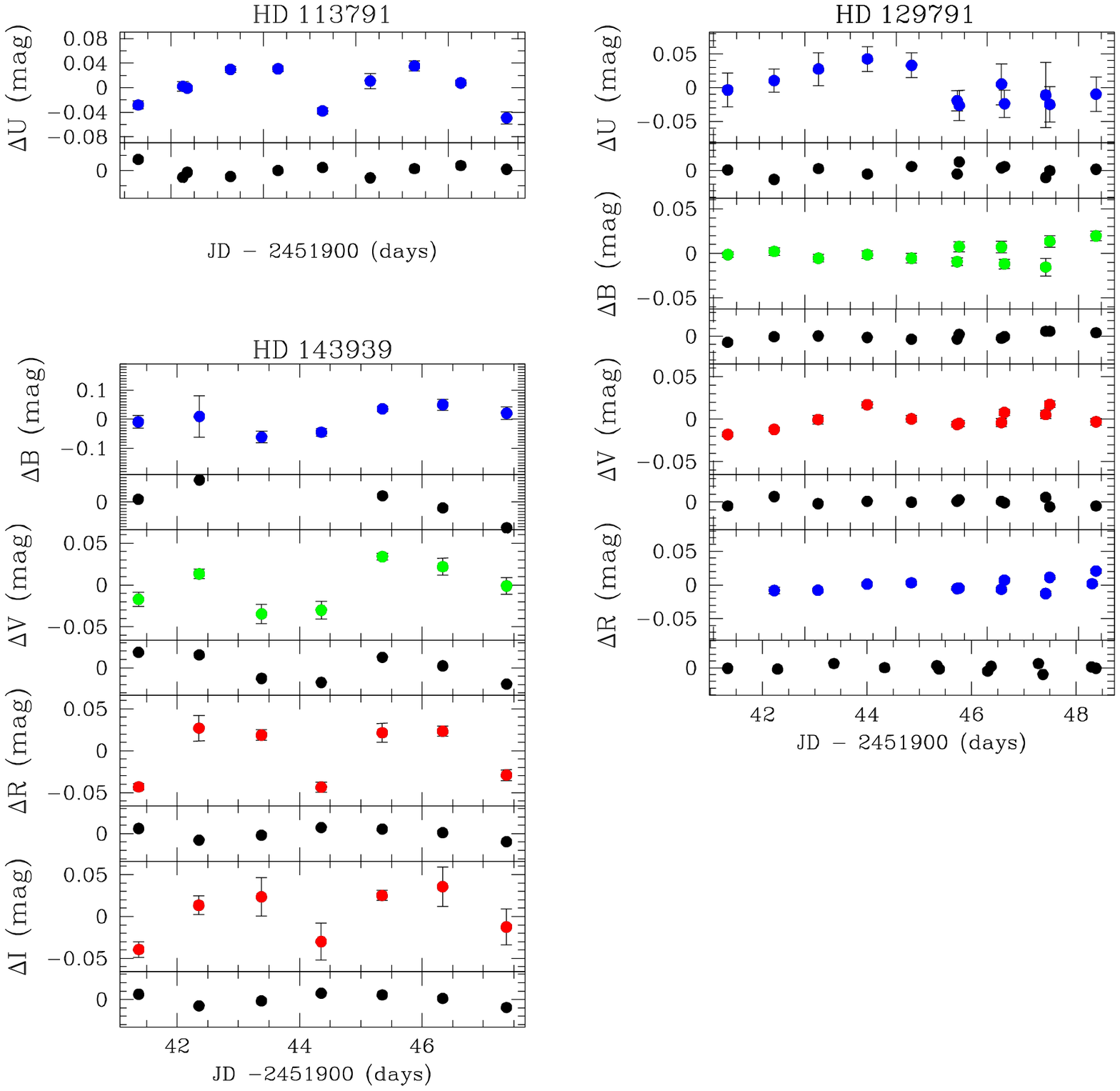}}
   \caption{cont.
}
\end{figure*}

The optical lightcurves of the sample in the BVR-bands are
displayed in Fig.  ~\ref{fig:light}. We have constructed them
subtracting the magnitudes of the targets from the magnitudes of an
'artificial comparison' star.  The 'artificial' stars have been
derived in the following way: for each observation we have analyzed
the lightcurves of several field stars in order to select those that
are not varying over the ten nights.  From these stars we have
rejected those displaying instrumental magnitudes and colors very
different from that of the targets. We have finally weighted the
selected stars according to their observational errors (the weight is
taken as the inverse of the square of the magnitude error,
$w=1/error^2$) and computed their averaged magnitude.  The
lightcurve of the corresponding 'artificial' star is shown in the panel 
below the lightcurve of each target in Fig.~\ref{fig:light}.

In order to detect variability in the stellar light of the Lindroos
stars, we have adopted the following criteria: we consider the
Lindroos stars to be variable when the standard deviation of their
mean magnitude over the ten nights is three times larger than the
standard deviation of the mean magnitude of the artificial star, that
is, $\sigma_{Lindroos\,PTTSs} \geq 3\,\sigma_{Artificial\,Star}$.

We have analyzed the lightcurves of each target in the five filters.
In general, the {\em U}-band data show a lower SNR when compared to
the {\em BVRI} filters given that the targets (and the comparison
stars) are fainter at that wavelength.  However, the amplitudes in
this band are larger, so in most of the cases we can use the {\em
U}-band to report significant variability of the targets.

\subsection{Determination of rotation periods}

The rotation periods have been computed from the photometric data
applying two different methods: the string-length (SL) method
proposed by Dworetsky (1983) and the Scargle periodogram (SP, Scargle
1982).

The string length method works in the phase space and it is well
suited for samples with a small number of unevenly spaced data points
(Dworetski 1983).  This method seems not to be well-suited for large and
clumped datasets (Scholz \& Eisl\"offel 2004).  In this method, a
rotational period is assumed and the phase-folded lightcurve is
constructed.  Afterward, the length of the string that joins
consecutive points in the folded diagram is computed. The
string-length algorithm is then applied to a set of trial periods and
the one showing the shortest string length is chosen as the best
period.  We have applied the algorithm to all the targets in the
different filters.

In order to check the significance level of the derived periods, we
have performed several tests. Firstly, we have generated 10000
synthetic lightcurves using a Monte-Carlo method.  The 10000 aleatory
samples have the same temporal sampling as our data (that is, we have
preserved the Julian Dates) but aleatory photometric values within the
amplitudes of the observational data.  We have applied the string
length method to the 10000 synthetic lightcurves.  Each data set
provides a minimum value of the string length that corresponds to the
best period.  The percentage of samples showing a string length larger
than the string length of the observational data provides the
confidence level of the period determination.  We assignee a 99\%
confidence level to a case in which all the randomized data show a
string length larger than that obtained with the observational data,
because we consider that we cannot discriminate between 99\% and
100\%. Whenever more than one period is found to be statistically
significant, the visual inspection of the data and the quality of the
phase-folded lightcurve has helped us to choose the most probable one.
The main disadvantage of this test is that if the random magnitudes
can adopt any value within the given amplitude, the probability of
having a string value larger than the one found in the real data can
be high. This translates into a high significance of the derived
period that may be overestimated.

As a second test, we have generated 10000 aleatory samples preserving
the Julian Date and randomly distributing the real magnitude values.
We have applied the SL method to the 10000 samples and kept the
minimum SL for each of them. The confidence level is given the
fraction of samples with SL values smaller than the one obtained with
the real data.  The results obtained with this method are similar to
the ones obtained with the first test.

As a final test, we have generated 10000 pseudo-aleatory samples
retaining the Julian Dates of the observations and generating
randomized magnitudes within a given interval.  The magnitudes are not
completely random in the sense that they can only adopt values
following this expression:

\begin{equation}
mag_{aleatory} = mag_{measured} \pm \Delta
\end{equation}

where $\Delta$ is a random generated increment that follows a Gaussian
distribution, with maximum value the upper limit for the photometric
error. The idea behind this method is to increase the noise of the
data and to check whether the period that we found in the original
data persists or not.  In general, we have adopted the derived
photometric error as the maximum value of the increment. However, we
have also used larger values to analyze the effect of a large noise in
our period search, given that a spurious period should be much
affected by the extra noise.  The fact that the increments follow a
gaussian function ensures that the probability of having a magnitude
that differs significantly from the measured one is smaller than the
probability of having a magnitude closer to the most probable
value. In this way, the string length of the random sample is not
artificially enlarged.  After applying the string length method to the
10000 aleatory samples generated in this way, we have obtained that
the previously derived periods persist. Moreover, we do not get
significant alias periods for most of the stars in the sample.

We have also analyzed the periodicity of our lightcurves using the
Scargle method, which works in the frequency domain.  We have used
Interactive Data Language (IDL) routines to derive the periodogram of
the sources within a period interval of P$_{min}$=1\,days to
P$_{max}$= 10\,days. In the case of HD\,60102, we have considered a
P$_{min}$=0.5\,days given the better temporal sampling of the data.
We have also derived the SP of the 'artificial' comparison stars and
we have checked that the periods obtained for the Lindroos sample are
not present in their periodograms.

Given that most of the lightcurves show unevenly distributed
datapoints, we have not applied Horne \& Baliunas (1986) to derive the
false alarm probability (fap).  Instead, we have generated 10000
aleatory samples preserving the Julian Date and randomly distributing
the magnitude values (as in the case of the SL method).  We have
derived the Scargle periodogram for each sample and retained the power
of the highest peak.  The significance level is given by the fraction
of samples with power values smaller than the one obtained with the
real data.

As an extra test, we have averaged the data points taken during a
single night and repeated the period search applying both the SL and
the SP methods. In the latter case we have derived the {\em fap}
following Horne \& Baliunas (1986). As a result, the derived periods
are similar to the ones derived previously.  The {\em fap}s are larger
than those derived with our significance method for some of the
stars. We think this result is related with the small number of data
points used to derive the periodogram.

The results provided by the two methods, SP and SL, are shown in
Table~\ref{table2}.  We can consider as more secure (or probable) periods
those that fulfilled the conditions: (i) the periods are similar in
both the V- and B-band.  We have chosen these two bands because they
provide the best combination of amplitude and signal to noise ratio.
Whenever possible, we have also used the R-band and/or the U-band
and/or the I-band data to confirm the periodicity; (ii) such
periodicity is not found in the comparison stars and (iii) the
confidence level is higher than 95\%.  We have classified as less
secure (or possible) those periods that only have data in one optical
filter but fulfilled statements (ii) and (iii) and those periods that
fulfilled (i) and (ii) but show confidence levels between 90-95\%.

The results for each star are discussed in detail in the next section.

\begin{figure*}
 \centering
 \resizebox{18cm}{!}{ 
 \includegraphics{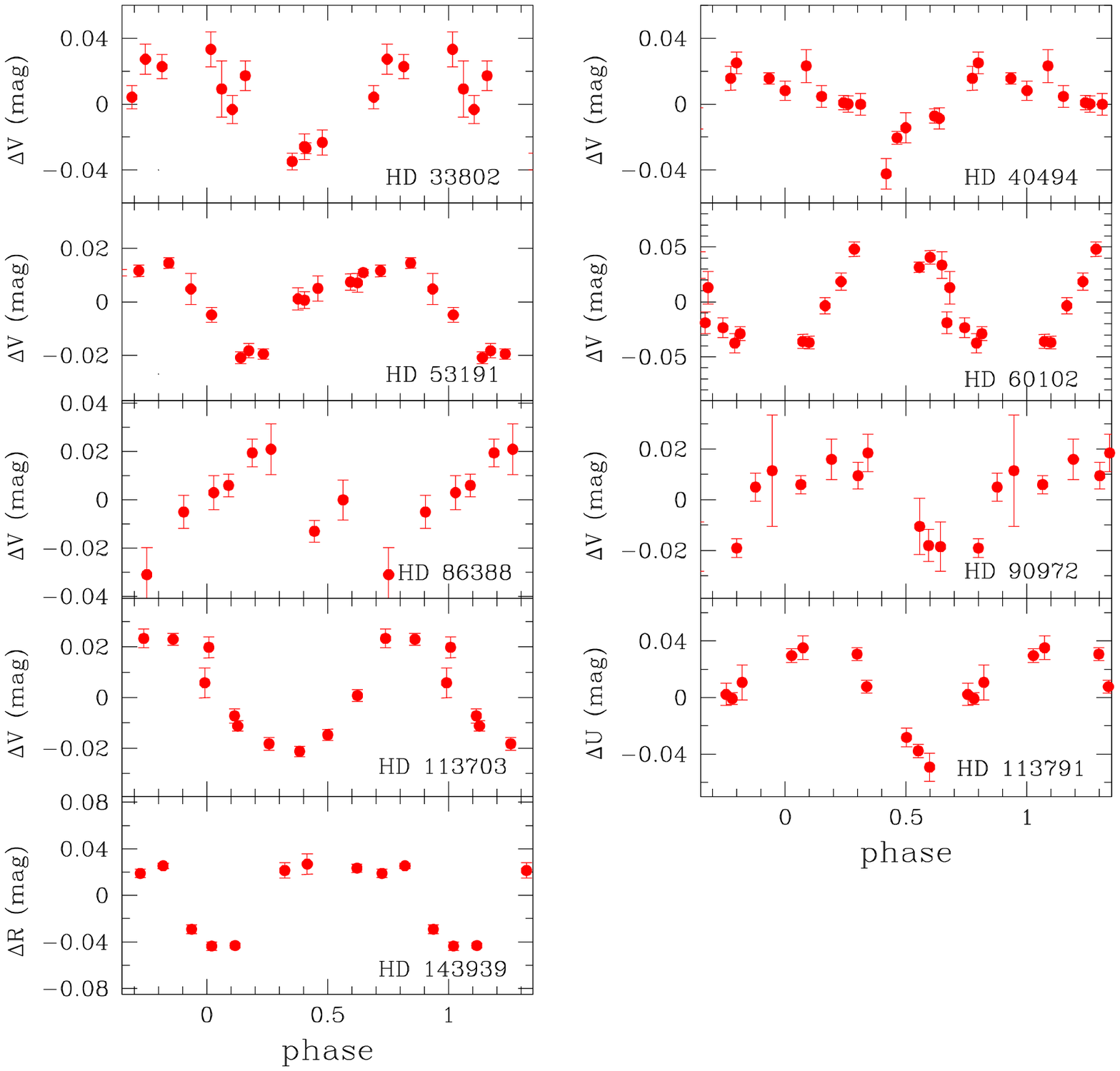}}
 \caption{V-band phase-folded lightcurves of the Lindroos PTTSs for
which we have found a rotational modulation.  The difference in
magnitude between the target and the 'artificial' star (see text) is
displayed against the Julian Date. The phases have been computed using
the rotation periods shown in Table~\ref{table2}.}
\label{fig:phasefolded}
\end{figure*}

\begin{table*}
\begin{center}
\caption{Photometric amplitudes in the {\em UBVRI} and rotational periods 
of PTTSs in Lindroos systems}\label{table2}
\begin{tabular}{llllllcccc}
\noalign{\smallskip}
          \hline          \noalign{\smallskip}
Star & $\Delta U$ & $\Delta B$ & $\Delta V$ & $\Delta R$ & $\Delta I$ &\multicolumn{4}{c}{period} 
 \\ 
 & & & &  & & SP (B-band) & SP (V-band) & SL (B-band) & SL (V-band) \\
 & (mag) & (mag) & (mag) & (mag) & (mag) & (days)& (days) & (days)& (days)   \\
 \noalign{\smallskip}
 \hline
 \noalign{\smallskip}
(a) & & &  & & &&  && \\\hline
HD\,33802\,B  & 0.15 & 0.08 & 0.07 & 0.05 &      & 2.8 (99) & 2.8 (99) & 2.8 (98) & 2.8 (99)\\
HD\,40494\,B  & 0.30 & 0.09 & 0.07 & 0.05 & -    & 6.1 (99) & 6.4 (99) & 6.4 (99) & 6.4 (99)\\
HD\,53191\,B  & 0.10 & 0.05 & 0.04 & 0.03 & 0.03 & 4.3 (99) & 4.4 (99) & 4.5 (97)& 4.4 (99)\\
HD\,60102\,B  & 0.14 & 0.12 & 0.09 & 0.09 & 0.09 & 1.8 (99) & 1.9 (99) & 1.8 (99) & 1.9 (99)\\ \hline
(b) & & & &  & & & & & \\ \hline
HD\,90972\,B  & -    & 0.05 & 0.04 & 0.04 & 0.04 & 3.2 (91) & 3.2 (91) & 3.4 (91)& 3.2 (93) \\
HD\,113703\,B & -    & 0.09 & 0.05 & 0.05 & 0.04 & 8.2 (99) & 7.7 (98) & 8.1 (99)   & 7.6 (99) \\
HD\,113791\,B & 0.08 & -    &-     &- &-  & 3.9 (96)$^1$ & - & 3.8$^1$ (98) & -   \\
HD\,143939\,B & -    & -    & -     & 0.07 & 0.07 & 3.0$^2$(97) & 3.2$^3$ (85)  & 3.0$^2$ (97)& 3.2$^3$ (97)\\     
\hline
(c) & & &  & & & & & & \\ 
\hline
HD\,86388\,B  & -    &-     & 0.05 & -    & -    & -   & 5.3 (85)& -   & 5.7 ($<$90) \\
HD\,38622\,C  & -    & 0.10 & 0.06 & 0.03 & -    & -   & -   & -   & -   \\
HD\,129791\,B & 0.07 & 0.03 & 0.03  & 0.03 & 0.02 & -       & - & -  & -    \\ 
    \hline
    \noalign{\smallskip}
\end{tabular}
\end{center}
 Notes: The sample has been divided in three groups according to the
uncertainties of the derived periods: (a) stars with more secure
(or probable) periods, (b) stars with less secure (or possible)
periods, and (c) stars without periods and stars for which we could
not derive a period; $^1$ Obtained with U-band data; $^2$ obtained
with R-band data; $^3$ obtained with I-band data
\end{table*}

\section{Results}

The phase-folded lightcurves of the targets have been plotted using
the periods provided in Table~2. They are shown in
Figure~\ref{fig:phasefolded}.  The amplitudes derived in the {\em BV}
bands are very similar to those derived by Wichmann et al. (1998) and
Bouvier et al.  (1997b) in their samples of PTTSs.  We have analyzed
the individual stars separately:

{\bf HD\,33802\,B:} it is a variable star according to its {\em BVR}
lightcurves (Figure~\ref{fig:light}). The data in the three filters
are in phase and they display a significant variation in comparison
with the artificial star. The derived period is 2.8 days for the {\em
BV} data, and a period of 3.0 days for the {\em R}-data. According to
the lightcurves, a period of 2.8 days fits better the observed
variation. We have obtained the same result with the two methods,
with a confidence level of 99\%.


{\bf HD\,38622\,C}: This PTTSs could not be monitored during the ten
consecutive nights given its proximity to the moon path. We have
plotted the obtained data points in Figure~\ref{fig:light}. The star
is variable in the {\em BVR} filters but we cannot report any
periodicity due to the scarcity of data.


{\bf HD\,40494\,B:} We have recorded a clear modulation of the light
in the {\em BVRI} filters.  The string-length method provides two
periods to fit the data: 6.4 days ({\em VB}-bands) and 5.6 ({\em
R}-band).  The minimum value of the string-length is obtained for 6.4
days with a significance of 99\%.  Moreover, the visual inspection of
the data suggests that a period larger than six days fits better the
observed modulation.  The SP method provides a period of 6.4\,d for
the V-band and 6.1\,d for the B-band. We have constructed the
phase-folded lightcurve using a value of 6.4\,d (see
Figure~\ref{fig:phasefolded}).


{\bf HD\,53191\,B:} The {\em UBVRI} lightcurves of HD\,53191\,B show a
clear modulation over the ten nights of observations. The minimum and
maximum values of the brightness are observed at the same Julian dates
in the five filters.  We have found a large number of comparison stars
in the field of view of HD\,53191\,B whose relative brightness is
constant over the ten nights.  The derived period is 4.4 days with the
two methods, with a significance level of 99\%.


{\bf HD\,60102\,B:} This source shows clear variability in all
lightcurves although it is difficult to see a periodic modulation in
the data. We observed the star three times during one of the nights
and we could see that the object is varying significantly in a short
timescale (see Figure~\ref{fig:light}).  We have derived a period of
1.9\,days ({\em UVR}-bands) that provides a smooth phase-folded
lightcurve.  The B-band provides a very close period of 1.8\,d.  We
obtain the same result with the two methods with a significance
level of 99\%.

If we averaged the data points, the SP method provides a period of
2.3\,days with a lower significance level ($\sim$96\% in {\em VR} and
smaller in the {\em UB}-bands).  The phase-folded lightcurve is not
well fitted with such period in any of the optical bands, so we have
rejected this period.


{\bf HD\,86388\,B} shows variations in the {\em V}-band lightcurve,
but we cannot study this variation in the {\em BRI} filters due the
lack of non-variable comparison stars.

We have applied the string-length method to the {\em V}-band data
obtaining a period of 5.7 days with a confidence level smaller than
90\%. When using the Scargle periodogram we derived a photometric
period of 5.3\,days and a confidence level of 85\%.  The visual
inspection of the V-band lightcurve shows a clear modulation. 
However, given that (i) there is only one band to derive the period
and (ii) the significance level is very low, we conclude that more
data is needed to validate the periodicity of this object.


{\bf HD\,90972\,B}: In the case of HD\,90972, the {\em U}-band data
are noisy and the artificial star shows a large scatter, so we have
only used the {\em BVRI} data to study the variability and periodicity
of the object.  The target shows a significant variability in these
four filters over the ten observing nights. The errors associated to
the data points are larger than in other stars because the separation
of this Lindroos system is small (11$\arcsec$) and the primary star is
very bright ($V$-mag = 5.6 mag). As a result, the multi-aperture
photometry is less accurate than in other targets.

The SL provides a period of $\sim$3.3\,d in {\em BVR}-bands with a
confidence level of 93\%. The SP provides a period of 3.2\,d (in
{\em BVR}-bands) with a 91\% confidence level. The observational
data are well fitted with a period of 3.3\,d, as shown in the phase-folded
lightcurve (Figure~\ref{fig:phasefolded}). However, we have classified
this period as 'possible' given its relatively low significance level.


{\bf HD\,113703\,B:} The PTTSs HD\,113703\,B was observed with the
1.5\,m Danish telescope on La Silla. We obtained lightcurves in the
{\em BVI} filters.  We have not considered the {\em R}-band data of
this object, given that it is too bright and it was saturated in some
of the exposures.  In the {\em U}-band there were no suitable
reference stars.

The {\em BV}-band lightcurves show a clear modulation and we have
found two significant periods: 7.6 days (obtained with the {\em
V}-band data and 8.0 days (obtained with the {\em B}- and {\em I}-band
data). We have also derived a shorter period of 0.9 days that fits the
observed data. We have found that this period is an alias of 8
days. Unfortunately, our time sampling is not short enough to confirm
a short period of 0.9\,days. The visual inspection of the lightcurves
suggests that a period of 8.0 days is the one that better reproduces
the observed modulation, so we have adopted this value as the most
probable. Given this discrepancy in the V and B-data and the
possibility of having a shorter period, we have classified the derived
period as 'possible'.  The phase-folded lightcurve is shown in
Figure~\ref{fig:phasefolded}.


{\bf HD\,113791\,B} could only be monitored in the {\em U}-filter
given that it is too bright at longer wavelengths, saturating the
detector.  The {\em U}-band lightcurve of HD\,113791\,B is displayed
in Figure \ref{fig:light}.  It shows a clear variability in its
brightness displaying a {\em U}-band amplitude 0.08 mag.  We have
derived a possible period of 3.8 days with a confidence level of 98\%.
A very similar result is obtained with the SP method.  As seen in
Figure~\ref{fig:phasefolded}, this rotation period fits the modulation
observed in the {\em U}-band data. Due to the lack of data in a
different optical filter, we have classified the period as
'possible'.

{\bf HD\,129791\,B}: this K-type star seems to be a variable source
according to its {\em U}-band lightcurve. The standard deviation of
the data is exactly 3 times that of the artificial star.  The
amplitude in this band is 0.07 mag, which is one of the smallest among
the sample.  In the {\em BVRI}-bands the amplitudes are $\sim$
0.03\,mag, with the artificial star varying at the same level, so
these bands do not provide further information.

There is no obvious modulation in the {\em U}-band lightcurve of
HD\,129791\,B. The star was observed twice during three different
nights, so we have studied if it could be a short-period variable.
After applying the SL and SP methods we do not find any significant
period.  Hence, we can only conclude that HD\,129791\,B seems to be a
variable star without an obvious periodicity within ten days of
monitoring.

{\bf HD\,143939\,B:} We have found non-variable comparison stars only
in the {\em R}- and {\em I}-band. Using the {\em R}-band data we have
derived a period of 3.0\,d which provides the phase-folded lightcurves
displayed in Figure~\ref{fig:phasefolded}. The confidence level is of
97\%.  We obtain the same result with both the SL and the SP methods.
In the case of the I-band we obtain a periodicity of 3.2\,d with a 
confidence level of 97\% when using the SL method. The SP method provides
a similar period value but a lower significance (85\%). Hence, we have
classified this period as 'possible'.

\section{Rotational properties of Lindroos PTTSs: comparison
with theoretical models}\label{sec:rotat-prop-lindr}

\begin{figure*}
\centering
 \resizebox{17cm}{!}{
 \includegraphics{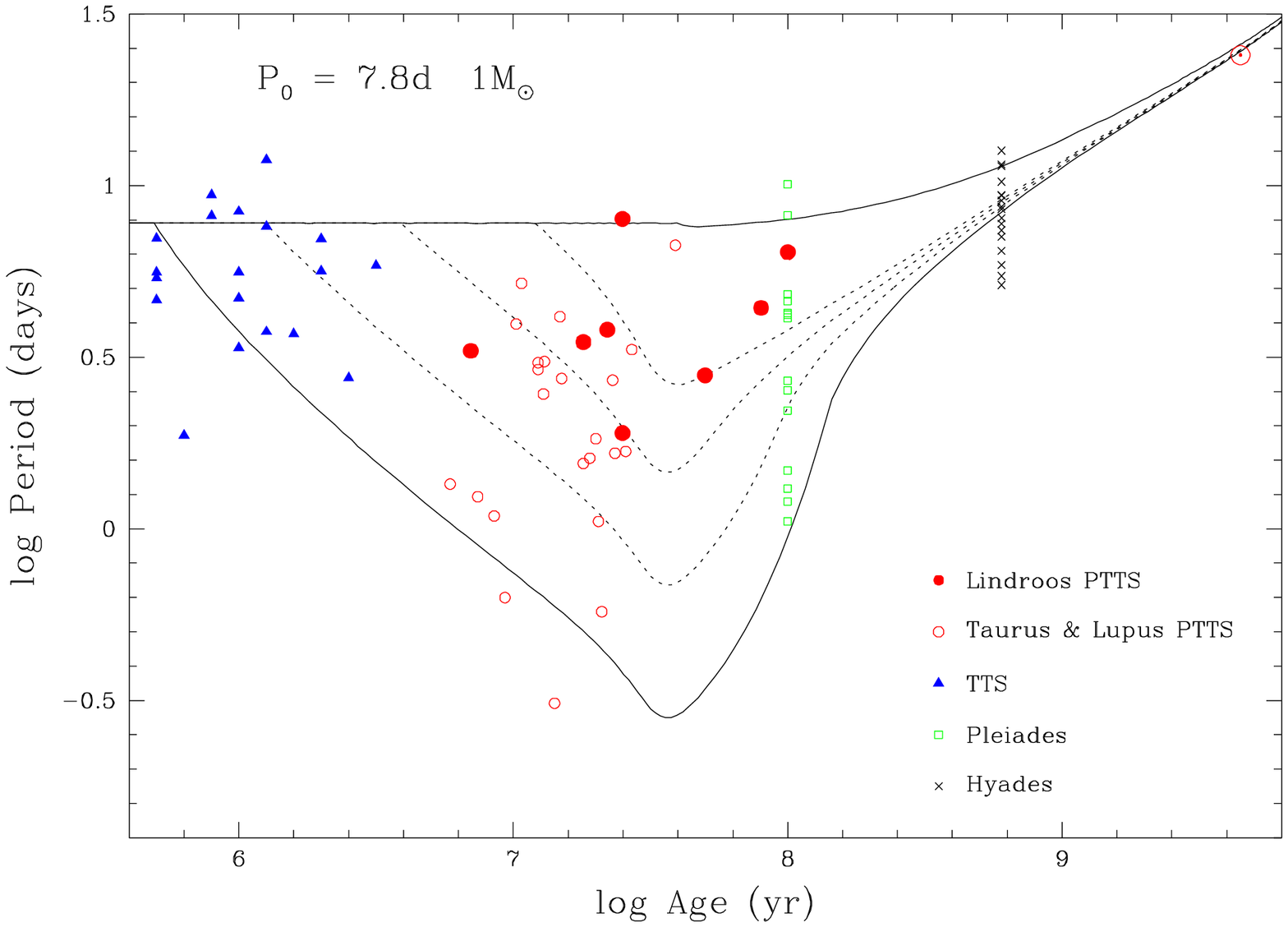}}
        \vspace{-3cm}
        \caption[Rotation evolutionary tracks for a 1\,$M_{\odot}$
        late-type star with an initial rotation period of
        $P_0\,(days)=7.8$: comparison with observational
        data]{Rotational evolution of a 1\,$M_{\odot}$ star with an
        initial rotation period of $P_0\,=\,7.8\,d$ .  The solid and
        dotted lines represent rotational tracks for different
        disk-star decoupling times ($\log\tau_{\rm disk}$ = 5.69,
        6.09, 6.59, 7.08 and 7.60 yrs).  The observational data
        correspond to the Lindroos PTTSs studied in this work (filled
        circles), the two samples of PTTSs candidates (open circles)
        studied by Bouvier et al.  (1997b) and Wichmann et al. (1998),
        TTSs in Taurus (filled triangles), and dwarfs stars in the
        Pleiades (open squares) and Hyades (crosses) clusters from
        Stelzer \& Neuh\"auser (2001).  Finally, the Sun is
        represented by an encircled dot.  The estimated masses of all
        late-type stars in the different groups range between
        0.9--1.2$M_{\odot}$.}
    \label{fig:bouvier1}
\end{figure*}

In order to examine the rotational properties of Lindroos PTTSs, we
have represented the sample with probable  and possible periods (groups
{\em a} and {\em b} in Table~\ref{table2}) in an age-rotation diagram
(Fig.~\ref{fig:bouvier1}).  The rotational evolution of late-type PMS
stars is mass-dependent. The estimated masses of the Lindroos PTTSs
with measured rotation periods range between 0.9--1.2 $M_{\odot}$ (see
Table~1), so we have focused our study on this mass range.

As shown in Fig.~\ref{fig:bouvier1}, the Lindroos PTTSs populate the
central part of the diagram with ages that range from 10-100\,Myr and
periods larger than 1.9\,d. We have represented the ages derived by
Palla \& Stahler (1999, PS99 hereafter) which results on a mixture of
PMS and ZAMS stars among the Lindroos sample.  If we consider the ages
provided by D'Antona \& Mazzitelli (1998, DM98), the Lindroos sample
populate a narrower age interval (10-35\,Myr) of the diagram (see
Table~1) and all the stars can be classified as PMS stars.

We have overplotted the PTTSs samples studied by Bouvier et
al. (1997b) and Wichmann et al. (1998).  In both cases we have only
considered stars with masses between 0.9 and 1.2\,$M_{\odot}$.  The
comparison of the three samples of PTTSs shows that they are
complementary: while most of the Lindroos PTTSs display periods longer
than 3\,d, the other two samples contain faster rotators that populate
the lower part of the diagram.  This result is probably related with
the different criteria used to select the samples: X-ray selected
samples mostly contains fast rotators, while non X-ray selected
samples can contain both slow and fast rotators. Our study has not
revealed the presence of very slow rotators among the Lindroos PTTSs,
that is, the slowest rotators among the Lindroos sample show rotation
periods comparable to the largest periods reported in Lupus and Taurus
PTTSs samples.

The rotational properties of the Lindroos systems can be compared with
the predictions of the disk-locking theory. In order to do this we
have used the rotational tracks by Bouvier et al. (1997a).
Fig.~\ref{fig:bouvier1} shows the theoretical age-rotation diagram for
1\,$M_{\odot}$ star (suitable to study the Lindroos PTTSs with masses
between 0.9-1.2$M_{\odot}$) and an initial rotation period of
$P_0=7.8$\,d for CTTSs. This value is based on the period distribution
of CTTSs in the Taurus SFR (Bouvier et al. 1993).  Younger late-type
stars presumably undergoing a mass accretion process (with masses $>$
0.25 M$_{\odot}$) in the Orion Nebula Cluster (ONC) show a broader
range of rotation periods but they also display a peak at 8\,d in
their period distribution. Only a 5\% of the sample of periodic
variables display periods longer than 12\,d (e.g. Attridge \& Herbst
1992, Eaton et al. 1995, Choi \& Herbst 1996, Stassun et al. 1999,
Herbst et al. 2002). Hence, we have kept the value of 8\,days as the
initial rotation period in the model.  The solid and dotted lines
represent the rotational tracks for different star-disk decoupling
times, which range between $5.69 \leq \log \tau_{\rm disk}\,(yr) \leq
7.60 $.

The Lindroos PTTSs lie in different regions within the diagram
(Fig.~\ref{fig:bouvier1}) which suggests different decoupling times
for the star-disk system.  Their rotational properties can be
explained assuming decoupling times between 1-20\,Myr.  Note that
HD\,113703\,B and HD\,113791\,B, the two PTTS close to its early
B-type primary, show relatively long rotation periods with estimated
decoupling times of 10 and 20\,Myr , respectively.  Both, the long
periods and the estimated decoupling times, suggest that the UV
radiation field of the early-type primaries have not played an
important role in the early dissipation of their disks.  When adopting
the ages provided by DM98 instead of PS99, we find a similar
decoupling time interval (10-20\,Myr) for the Lindroos sample.
However, three stars displaying younger ages (HD\,33802\,B,
HD\,40494\,B and HD\,53191\,B) show different properties: although
they remain almost on the same rotational tracks, they lie at the end
of the spin-up phase, while PS99 ages suggest that the three of them
are spinning down.  

All our conclusions are model-dependent, given the uncertainties
in the initial rotation periods of the Lindroos sample.  If we assume
a longer initial rotation period (15\,d) in the theoretical model, the
absolute values of the star-disk decoupling times are different for
the individual Lindroos PTTS. However, their rotational properties can still be
explained with decoupling times between 1-20\,Myr.  In the case of
shorter initial periods (e.g. 4.4 \,d), all the decoupling times are
longer than 5\,Myr while the stars with longer periods are not
well-fitted by the model.

If we include the other two samples of PTTSs (Taurus and Lupus) in
our analysis, a slightly different range of decoupling times is
necessary to explain their rotational properties ($\sim$0.5-10\,Myr).
And, finally, if we consider other samples of late-type stars
at different evolutionary stages (TTS, PTTS and ZAMS stars) a broader
range of decoupling times (0.5-20\,Myr) is required to explain their
position in the age-rotation diagram (see Figure~\ref{fig:bouvier1}).

The decoupling times are related to the lifetime of optically thick
circumstellar disks.  According to Figure~\ref{fig:bouvier1}, the
existence of both short and long-lived disks is required to explain
the different rotational properties of Lindroos, Taurus and Lupus PTTS
stars.  Infrared studies of young stellar clusters have shown that the
mean lifetime of protoplanetary disks is $\sim$ 6\,Myr (Haisch et
al. 2001).  This disk lifetime can explain the rotational properties
of most of the TTSs and part of the PTTSs and ZAMS stars represented
in Figure~\ref{fig:bouvier1}.  However, it cannot explain the presence
of slow rotators among the PTTSs and Pleiades samples.

The existence of long-lived optically thick disks has been studied by
Strom (1995) who showed that 30\% of 10\,Myr old late-type PMS stars
in the L\,1641 cloud display infrared excesses.  More recently, Lyo et
al. (2003) have reported the presence of circumstellar disks around
60\% of 10\,Myr old late-type stars in the young cluster $\eta$
Chamaleontis.  One important result from their study is that 30\% of
the stars with infrared excesses also show on-going accretion
processes.  This result supports the possibility of having decoupling
times of $\sim$ 10\,Myr among late-type PMS stars and could explain
the presence of 10-100 Myr old slow rotators in the age-rotation
diagram.  The Lindroos stars located in this part of the diagram have
already decoupled from their disks which have presumably dissipated at
earlier stages in their evolution.  Hence, we do not expect these
PTTSs to show IR excesses at near-IR wavelengths but maybe at longer
wavelengths as a result of grain growth during disk evolution
(Brandner et al. 2000).  In fact, some Lindroos PTTS were observed by
the Infrared Space Observatory (ISO) at 6.5 and 15 $\mu$m and display
mid-IR excesses. This result is interpreted as evidence of remnant
circumstellar matter (Moneti et al. 1998).

\section{The age-activity-rotation relation in PTTSs}\label{sec:aar}

The connection between coronal activity and rotation in late-type PMS
stars has been extensively studied in different works
(e.g. Pallavicini et al.  1981, Bouvier 1990, Stauffer et al. 1994,
Neuh\"auser et al. 1995, Flaccomio et al. 2003). These studies have
revealed that most 1\,M$_{\odot}$ late-type PMS stars with ages less
than 10 Myr display a high level of magnetic activity.  Stelzer \&
Neuh\"auser (2001) have studied the rotation-age-activity relation for
TTSs in Taurus (with ages between 10$^5$--10$^7$ yrs) in comparison to
dwarf stars in the Pleiades (age$\sim$ 10$^8$ yrs; Meynet et al. 1993)
and the Hyades (age$\sim$6$\times$10$^{8}$ Myr; Perryman et al. 1998).
They have reported a tight connection between coronal activity and
rotation for all three groups.

Lindroos PTTSs are intermediate between TTSs and Pleiades, so we have
studied the rotation-activity relations for our sample and compared it
with other groups of younger and older late-type stars.  

\subsection{X-ray luminosity versus rotation period}

\begin{figure*}
\centering \resizebox{16cm}{!}
 {\includegraphics{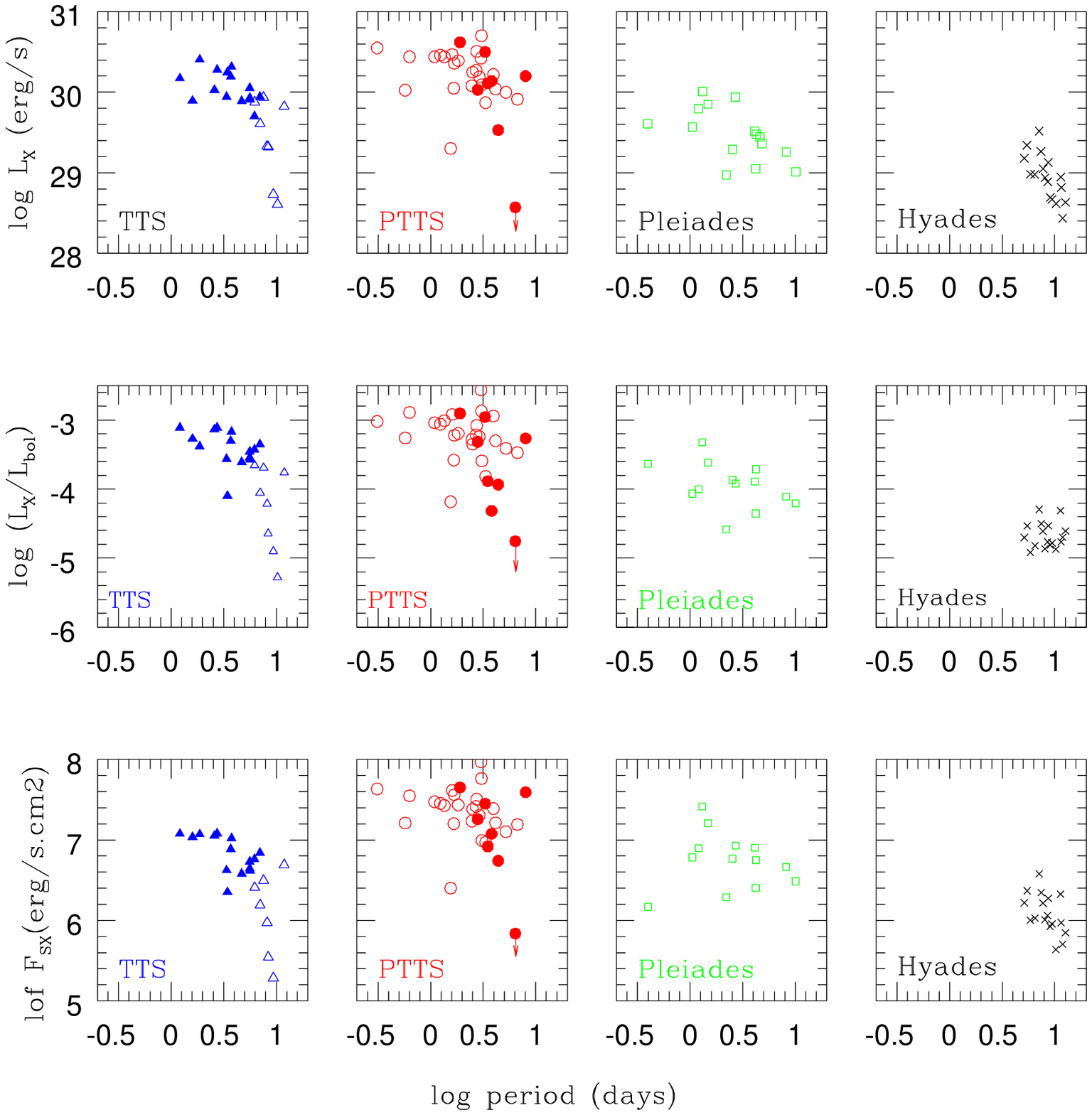}}
 \caption{Activity-rotation relations for late-type PMS and ZAMS
 stars with masses between 0.9-1.2\,M$_{\odot}$. We have represented
 TTSs in Taurus (CTTSs: open triangles, WTTSs: filled triangles),
 PTTSs (Lindroos: filled circles, Lupus \& Taurus: open circles),
 Pleiades and Hyades stars.  We have plotted three different activity
 indicators versus the rotation period: (i) X-ray luminosity (ii)
 X-ray to bolometric luminosity ratio and (ii) X-ray surface flux.
 The relations are very similar for PTTSs and TTSs. Both groups
 contain very active stars with $\log (L_{\rm X}/L_{\rm bol})$ ratios
 close to the saturation level of --3.}
 \label{fig:activity}
\end{figure*}

We have started our study comparing the rotation period and the X-ray
luminosity of the Lindroos PTTSs.  As seen in the upper panel of
Fig.~\ref{fig:activity}, Lindroos PTTS show a clear anti-correlation
between rotation and X-ray emission being the strongest X-ray emitters
are the fastest rotators.  The other two samples of PTTS contain
faster rotators. They show a flatter relation with a larger spread in
rotation periods for a given X-ray luminosity.

In order to analyze the evolution of the $\log L_{\rm X}$-period
relation with age, we have also included in Fig.~\ref{fig:activity}
three samples of late-type stars at different evolutionary stages:
TTSs in Taurus, Pleiades and Hyades dwarf stars.  The observational
data for these three samples have been taken from Stelzer \&
Neuh\"auser (2001) and references therein.  As in the previous
section, our study is focused on late-type stars with masses between
0.9--1.2$M_{\odot}$.

PTTSs and TTSs show a very similar X-ray luminosity-period relation:
the strongest X-ray emitters display a broad range of rotation periods
(from 0.3\,d-10\,d) and show no correlation between X-ray luminosity and
rotation period.  On the other hand, the less active stars show
periods longer than $\sim$4\,d and display a steep relation between
X-ray luminosity and rotation with the strongest X-ray emitters being
the fastest rotators.

Five Lindroos PTTSs occupy the same region of the diagram as WTTSs,
displaying the strongest X-ray luminosity of the whole sample.
Their X-ray emission is not correlated with the rotation
period.  On the other hand, two Lindroos systems (HD\,40494\,B,
HD\,53191) display lower activity levels.  These are the oldest stars
in the sample.  According to PS99, HD\,40494 and HD\,53191\,B show
ages of 80\,Myr and 100\,Myr respectively, so they could be
considered as ZAMS stars. In that case, their position in the diagram
could reflect an age effect, that is, they show lower levels of X-ray
emission due to their older ages. However, if we consider DM98 tracks,
these stars are significantly younger (35\,Myr and 30\,Myr,
respectively) and an age effect is not longer able to explain their
position in the diagram. In this case, they could be considered the
less active PTTSs among the sample.

According to Fig.~\ref{fig:activity}, the most active late-type PMS
stars ($\lg\,L_{\rm X} \geq 30\,$erg/s) do not show a correlation
between activity and rotation: for a given X-ray luminosity we found a
wide range of rotation periods. On the other hand, the weakest X-ray
PMS emitters $\lg\,L_{\rm X} < 30\,$erg/s) display a correlation between
the X-ray luminosity and rotation period. At the age of the Pleiades,
all stars display X-ray luminosities $\lg\,L_{\rm X} \leq
30$~erg/s~and show a connection between activity and rotation.

The comparison of PTTS with Pleiades and Hyades stars shows a
connection between activity-rotation with age.  ZAMS stars do not
display a $\log L_{\rm X}$-period relation as steep as that found for 
less active late-type PMS stars. Moreover, the fastest rotators in the
Pleiades exhibit lower X-ray luminosities than those from PTTSs and
TTSs samples.  The decrease in the magnetic activity is more evident
in the Hyades, where the mean X-ray luminosity is one order of
magnitude smaller than that from the other three samples.

\subsection{ $(L_{\rm X}/L_{\rm bol})$ ratio and X-ray 
surface flux versus rotation period}

The X-ray luminosity depends, among other factors, on the stellar
radius. We have taken into account this effect studying normalized
activity parameters like the ratio between the X-ray and the
bolometric luminosity, $L_{\rm X}/L_{\rm bol}$, and the X-ray surface
flux, $F_{\rm SX}$, of the Lindroos PTTSs.  We have computed these
parameters adopting the bolometric luminosities and effective
temperatures listed in Gerbaldi et al. (2001).

The $\log(L_{\rm X}/L_{\rm bol})$ ratio has been derived for a large
number of late-type star samples.  In general, the most active stars
show ratios close to the saturation value of $\log(L_{\rm X}/L_{\rm
bol})=-3$ (e.g. Fleming et al. 1989, Feigelson et al.  1993, Micela et
al. 1999): late-type stars displaying that ratio have reached the
highest level of X-ray activity, e.g.  by coverage of the whole
surface with magnetic spots.  The middle panel of
Fig.~\ref{fig:activity} shows the relation between the $\log(L_{\rm
X}/L_{\rm bol})$ ratio and the rotation period for TTSs, PTTSs,
Pleiades and Hyades stars.  In the case of the Lindroos PTTSs, the
relation is very steep and includes both saturated and unsaturated
stars.  Three of the stars display ratios close to the saturation
value of --3 and short rotation periods.

  The Lindroos sample display a $\log(L_{\rm X}/L_{\rm bol})$-period
relation very similar to that found in TTSs, containing the two
samples very active stars with both short and long rotation
periods. The most active PTTSs show a relation very similar to that of
WTTSs while the less active Lindroos PTTSs occupy the same region in
the diagram as CTTSs.  In both groups, TTSs and PTTSs, we find stars
with periods between 1--8 days and with $\log(L_{\rm X}/L_{\rm bol})$
ratios close to the saturation value of --3, indicating that during
the PMS phase the saturation level is reached with different
rotational properties.  ZAMS and MS stars show flatter
activity-rotation relations, with all of them displaying $\log(L_{\rm
X}/L_{\rm bol})$ ratios lower than the saturation value of --3.  In the
case of the Hyades, most of the targets exhibit $\log(L_{\rm X}/L_{\rm
bol})$ ratios close to the value of --4.5 and long periods.

  Flaccomio et al. (2003) have shown that at ages between 1-7 Myr most
of the 1\,$M_{\odot}$ late-type PMS stars are saturated and their
X-ray and rotational properties are not correlated.  They report a
decrease in the magnetic activity at the age of the Pleiades
($\sim$100\,Myr) suggesting that this decrease must take place at ages
older than 10\,Myr.  We find a mixture of saturated and non-saturated
stars among the Lindroos sample. This result suggests high activity
levels at ages of 10\,Myr and a decrease in coronal activity at longer
ages.

Finally, we have studied the relation between the rotation period and
the X-ray surface flux for PTTSs, TTSs, Pleiades and Hyades stars
(lower panel in Figure~\ref{fig:activity}). The X-ray surface flux is
implicitly related to the fraction of stellar surface covered with
active regions.  As seen in Fig.~\ref{fig:activity}, there is a clear
variation in the steepness of the relation from the younger stars to
the older ones. As in the case of the $\log(L_{\rm X}/L_{\rm bol})$
ratio, TTS and PTTS display very similar relations although the most
active stars among the PTTSs sample show larger X-ray surface fluxes
than the most active TTSs.  These could be related to the fact that
TTSs are still contracting to the MS, displaying larger surface areas
than PTTSs. The latest have finished their contraction and approach
the MS following the radiative part of the evolutionary tracks.  The
$\lg\,F_{\rm SX}$-period relation becomes flatter for stars in the
Pleiades and the Hyades.

\section{Conclusions}\label{sec:conclusions}

In this preliminary study, we have analyzed the rotational
properties of a sample of PTTSs in Lindroos systems.  We have derived
their rotation periods after an optical monitoring of the targets
during two campaigns of 10 nights each.  We have compared the derived
periods with theoretical PMS rotational tracks.  Finally, we have
studied the activity-rotation relation for PTTSs and compared it with
samples of younger and older stars.  The main results of this study
can be summarized as follows:

{\bf 1.} The optical monitoring of the Lindroos sample has allowed us
to study the rotational properties of 11 PTTSs in Lindroos
systems. The analysis of the optical data revealed that most of the
Lindroos PTTSs are variable. The amplitudes computed in both the {\em
B}- and {\em V}-band are comparable to those measured in other samples
of PTTSs candidates.  Our analysis has shown that eight (out of
11) Lindroos PTTS show periodic modulations in their lightcurves.  We
have classified the derived periods as probable or possible
depending on their confidence levels.  Finally, we could not report
any periodicity for these three targets: HD\,38622, HD\,86388 and
HD\,129791.  This preliminary work complements previous
rotational studies of PTTSs candidates in which the samples were X-ray
selected and, therefore, biased towards fast rotators.

{\bf 2.}  We have examined the rotational properties of the Lindroos
sample in comparison with two other samples of PTTSs through a
period-age diagram: while most of the Lindroos PTTSs have rotation
periods longer than 3 days, this is not the case of the other two
samples that mostly contain fast rotators. We think this is related
to the different sample selection criteria.  The three samples of
PTTSs fill the region between 10$^7$--10$^8$~yrs, displaying a broad
range of periods that range between 0.3 and 8.0\,d.  It is indeed at
such age interval that theoretical models predict the largest
dispersion of the rotational properties of late-type PMS stars, as a
result of the spin-up of late-type PMS stars in their approach to the
MS.

We have compared the rotation periods of Lindroos PTTS with
theoretical models by Bouvier et al. (1997a).  The largest scatter in
rotation periods corresponds to the PTTSs phase, and it can be
explained as a result of the different decoupling times between the
star-disk system during the PMS phase.  The rotational properties of
the Lindroos PTTSs can be explained assuming different star-disk
decoupling times, between $\sim$ 1-20 Myr, for an initial period of
8\,d.  If we consider other samples of PTTSs, the range of decoupling
times is broader (0.5-20\,Myr).  This range of times is directly
related to the lifetime of circumstellar disks.  IR studies of young
stellar clusters have shown that disks can survive up to 10\,Myr.  The
reason why disks could dissipate at such different timescales is still
an open question, although the initial star formation conditions (dust
and gas content in the initial cloud) and the environment in which
disks evolve (close to early-type stars) could certainly have an
influence in their lifetimes.

{\bf 3.}  We have studied the connection between magnetic activity and
rotation for Lindroos PTTSs through three different activity
indicators: the X-ray luminosity ($L_{\rm X}$), the X-ray to the
bolometric luminosity ratio $(L_{\rm X}/L_{\rm bol})$ and the X-ray
surface flux ($F_{\rm SX}$).  We have reported steep activity-rotation
relations for Lindroos PTTSs.  In general, the less active stars
display longer rotation periods.  Some of the Lindroos PTTSs are very
active, displaying $(L_{\rm X}/L_{\rm bol})$ ratios close to the
saturation value of --3. This could be the result of the spin-up
phase.  The fact that the Lindroos sample also contains non-saturated
stars with low levels of magnetic activity could indicate that part of
the sample has started to spin-down.  This result is consistent with 
a decrease in the coronal activity at ages $>$10\,Myr, as  suggested by
Flaccomio et al. (2003).

{\bf 4.} We have compared the activity-rotation relations of PTTSs
with those displayed by different samples of younger and older
late-type stars: TTSs in Taurus, Pleiades and Hyades dwarfs stars.
PTTSs show rotation-activity relations very similar to those from TTSs
mainly because both groups contain young and magnetically active
stars. Some of the stars display $\log(L_{\rm X}/L_{\rm bol})$ ratios
closer to the saturation limit.  The rotation periods of these stars
range between 1--8\,d, that is, PMS late-type stars reach the
saturation level with different rotation properties.  When comparing
PMS stars with ZAMS and MS stars, we see that older stars display
$\log (L_{\rm X}/L_{\rm bol})$ ratios smaller than the saturation
limit of --3.

Future monitoring of a larger sample of PTTS for a longer period of
time will be important to confirm the conclusions derived in this
preliminary study.

\begin{acknowledgements}
We thank the referee, W. Herbst, for useful comments that 
improved the manuscript.  We thank CTIO and La Silla staff, in special
E. Cosgrove and F. Selman, for their support during the observations.
NH thanks B. Stelzer and A. Scholz for useful comments on the
manuscript and M. Billeres for her assistance. MF was partially
supported by the Spanish grant AYA2001-1696. NH and RN did most of
this work at MPE Garching, where RN was supported by the BMBF
Verbundforschung ROSAT.

\end{acknowledgements}


\begin{thebibliography}{}

\bibitem{} Allain S., 1998, A\&A 333, 629

\bibitem{} Allain S., Bouvier, J., Prosser, C., et al., 1996, A\&A, 305, 498

\bibitem{} Attridge J.M. \& Herbst W., 1992, ApJ 398, L61 
 
\bibitem{} Bouvier J. \& Corporon P., 2001, {\em The formation 
   of binary systems}, IAU Symp. 200, ed. H. Zinnecker and R.D. Mathieu  

\bibitem{} Bouvier, J., Forestini, M. and Allain, S., 1997a, A\&A 326, 1023 

\bibitem{} Bouvier, J., Wichmann, R., Grankin et al. 1997b, A\&A 318, 495 

\bibitem{} Bouvier, J., Cabrit, S., Fern\'andez, M. et al., 1993, A\&A 272, 176

\bibitem{} Bouvier, J., 1990, AJ 99, 946 

\bibitem{} Brandner, W., Zinnecker, H., Alcal\'a, J. et al., 2000, AJ
120, 950

\bibitem{} Choi, P.I. \& Herbst, W., 1996, AJ 111, 283

\bibitem{} Colllier Cameron, A., \& Campbell, C.G., 1993, A\&A 274, 309

\bibitem{} D'Antona, F. \& Mazzitelli, I., 1998, ApJS 90, 467, (DM98)

\bibitem{} Duquennoy, \& Mayor, M., 1991, A\&A 248, 485

\bibitem{} Dworetsky, M.M., 1983, MNRAS 203, 917

\bibitem{} Eaton, N.L.,  Herbst, W. and Hillenbrand, L.A. , 1995, 
    AJ 110, 1735

\bibitem{} Edwards, S., 1993, Strom, S.E., Hartigan, P., et al., 
1993, AJ 106, 372

\bibitem{} Feigelson, E.D., Casanova, S., Montmerle, T. et al., 1993,
         ApJ 416, 623

\bibitem{} Flaccomio, E., Micela, G. and Sciortino, S., 2003, A\&A 402, 277

\bibitem{} Fleming, T.A., Gioia, I.M. and Maccacaro, T., 1989, ApJ 340, 1011

\bibitem{} Gerbaldi, M., Faraggiana, R. and Balin, N., 2001, A\&A 379, 162

\bibitem{} Haisch, K., Lada, E.A. and Lada, C.J., 2001, ApJL 553, 153

\bibitem{} Herbig, G.H., 1978, {\em Problems of Physics and evolution of
the Universe}, Eds. L.V. Mirzoyan, Publishing House of the Armenian
Academy of Sciences, p. 171

\bibitem{} Herbst, W., Maley, J.A. and Williams, E.C., 2001a, AJ 120, 349

\bibitem{} Herbst, W., Bailer-Jones, C.A.L. and Mundt, R.,  2001b, 
ApJL 554, 197

\bibitem{} Herbst, W., Bailer-Jones, C.A.L., Mundt, R. et al., 2002, 
A\&A 396, 513

\bibitem{} Horne J.H.  \& Baliunas S.L. (1986), ApJ ,302, 757
 
\bibitem{} Hu\'elamo, N., Neuh\"auser, R., Stelzer, B., Supper, R.
 and Zinnecker, H., 2000, A\&A 359, 227

\bibitem{} Jensen E.L.N., 2002, {\em Young Stars Near Earth: Progress
and Prospects}, ed. R. Jayawardhana \& T. Greene, ASP Conference
Series, vol. 244, p.3


\bibitem{} Johnstone, D., Hollenbach, D. and Bally, J., 1998, ApJ 499, 758

\bibitem{} K\"onigl, A., 1991, ApJ 370, L39 

\bibitem{} Lamm, M., Mundt, R., Bailer-Jones, C.A.L. et al., 2004, A\&A, 
417, 557


\bibitem{} Lindroos, K.P., 1985, A\&AS 60, 183

\bibitem{} Lindroos, K.P., 1986, A\&A 156, 223

\bibitem{} Lyo, A., Lawson, W.A., Mamajek, E.E. et al., 2003, MNRAS 338, 616

\bibitem{} Mart\'{\i}n, E.L., Magazz\`u, A., Rebolo, R., 1992, A\&A 257, 186

\bibitem{} Mathieu, R.D., 1992, {\em Binaries as tracers of stellar
   formation}, ed. Duquennoy \& Mayor, Cambridge University Press, p.155

\bibitem{} Melo, C.H.F., Covino E., Alcal\'a J.M. and Torres G., 2001, 
  A\&A 378, 898

\bibitem{} Meynet, G., Mermilliod J.C. and Maeder, 1993, A\&AS 98, 447

\bibitem{} Micela, G., Sciortino, S., Harnden, Jr.F.R. et al., 1999, 
    A\&A 341, 751

\bibitem{} Moneti, A., Zinnecker, H., Kunkel, M. 
      and Preisbich, T., 1998, Proc. {\it Star formation
      with the Infrared Space Observatory}, eds. J. L. Yun \& R. Liseau

\bibitem{} Neuh\"auser, R., Sterzik, M.F., Schmitt, J.H.M.M.,
 Wichmann, R. and Krautter, J., 1995, A\&A 297, 391

\bibitem{} Palla, F. \& Stahler, S.W., 1999, ApJ 525, 772, (PS99)

\bibitem{} Pallavicini, R., Golub, L., Rosner, R., et al., 1981, 
ApJ 248, 279
 
\bibitem{} Pallavicini, R., Pasquini, L. and Randich, S., 1992, A\&A 261, 245

\bibitem{} Pan, K., Tan, H. and Shan, H., 1998, A\&A 335, 179

\bibitem{} Perryman, M.A.G., Brown, A.G.A., Lebreton, Y. et al., 1998, 
        A\&A 331, 81

\bibitem{} Prosser, C.F., Shetrone, M.D., Dasgupta, A., et al., 1995, 
PASP 107, 211

\bibitem{} Queloz, D., Allain, S., Mermilliod, J.C., et al., 
1998, A\&A, 335, 183

\bibitem{} Radick, R.R., Thompson, D.T., Lockwood, G.W., et al., 
1987, ApJ, 321, 459

\bibitem{} Rebull, L.M., 2001, AJ 121, 1676

\bibitem{} Rebull, L.M., Wolff, S.C., Strom, S.E. 
and Makidon, R.B., 2002, AJ 124, 546

\bibitem{} Scargle, J. D., 1982, ApJ 263, 835
 
\bibitem{} Scholz A. \& Eisl\"offel J., 2004, A\&A, 421, 259 

\bibitem{} Shu, F., Najita, J. Ostriker, E., et al., 1994, ApJ 429, 781

\bibitem{} Siess, L., Dufour, E. \& Forestini, M., 2000, A\&A 358, 593

\bibitem{} Soderblom, D.R., Stauffer J.R., MacGregor, K.B, and Jones B.F., 1993, 
 ApJ 409, 624

\bibitem{} Stassun, K., Mathieu, R.D., Mazeh, T. and Vrba, F.J., 
 1999, AJ 117, 2941

\bibitem{} Stauffer, J.R., Balachandran, S.C., Krishnamurthi, 
A. et al., 1997, ApJ 475, 604

\bibitem{} Stauffer, J.R., Caillault, J.P., Gagn\'e, M. et al., 1994,
1994, ApJS 91, 625

\bibitem{} Stauffer, J.R, Hartmann, L., Jones, B.F., 1989, ApJ 346, 160

\bibitem{} Stelzer, B. and Neuh\"auser, R., 2001, A\&A 377, 538

\bibitem{} St\"orzer, H. \& Hollenbach D., 1999, ApJ 515, 669

\bibitem{} Strom, S., 1995, RMxAC 1, 317  

\bibitem{} Terndrup, D., Stauffer, J.R., Pinsonneault, M.H. 
et al.,  2000, AJ\,119, 1303 

\bibitem{} Tout, C.A., Livio, M. \& Bonnell, I. A., 1999, MNRAS 310, 360

\bibitem{} Webb, R.A, Zuckermann, B., Platais, I. et al., 1999,
ApJ 512, L63

\bibitem{} Wichmann, R., Bouvier, J., Allain, S. et al., 1998
   A\&A 350, 521

\bibitem{} Zahn J.P., 1977, A\&A 57, 383


\end{thebibliography}
\end{document}